\documentclass[twocolumn]{aastex63}
\usepackage[utf8]{inputenc}
\pdfoutput=1
\usepackage[english]{babel}
\usepackage{graphicx}
\usepackage{color,soul}
\usepackage{comment}
\graphicspath{{.}}
\usepackage{lineno}

\received{-}
\revised{-}
\accepted{-}

\submitjournal{PSJ}

\begin{document}

\newcommand{\jdlong}{2015~JD$_{1}$} 

\newcommand{\jd}{JD$_{1}$}

\title{Physical Characterization of \jdlong: A Possibly Inhomogeneous Near-Earth Asteroid}

\correspondingauthor{Andy J. L\'{o}pez-Oquendo}
\email{al2987@nau.edu}

\author[0000-0002-2601-6954]{Andy L\'{o}pez-Oquendo}
\affiliation{Department of Astronomy and Planetary Science, Northern Arizona University, Flagstaff, AZ 86011, USA}

\author[0000-0003-4580-3790]{David E. Trilling}
\affiliation{Department of Astronomy and Planetary Science, Northern Arizona University, Flagstaff, AZ 86011, USA}

\author[0000-0002-7600-4652]{Annika Gustafsson}
\affiliation{Department of Astronomy and Planetary Science, Northern Arizona University, Flagstaff, AZ 86011, USA}

\author[0000-0002-4129-5381]{Anne Virkki}
\affiliation{Arecibo Observatory, University of Central Florida, Arecibo, PR 00612, USA}
\affiliation{Department of Physics, P.O. Box 64, FI-00014 University of Helsinki, Finland}

\author[0000-0002-4042-003X]{Edgard G. Rivera-Valent\'{i}n}
\affiliation{Lunar and Planetary Institute, Universities Space Research Association, Houston, TX 77058, USA}

\author[0000-0002-5624-1888]{Mikael Granvik}
\affiliation{Department of Physics, P.O. Box 64, FI-00014 University of Helsinki, Finland}
\affiliation{Asteroid Engineering Lab, Lule\aa{} University of Technology, Box 848, S-981 28 Kiruna, Sweden}

\author[0000-0001-7335-1715]{Colin Orion Chandler}
\affiliation{Department of Astronomy and Planetary Science, Northern Arizona University, Flagstaff, AZ 86011, USA}

\author[0000-0002-1278-5998]{Joseph Chatelain}
\affiliation{Las Cumbres Observatory, 6740 Cortona Drive Suite 102, Goleta, CA 93117, USA}

\author[0000-0002-2493-943X]{Patrick Taylor} 
\affiliation{Lunar and Planetary Institute, Universities Space Research Association, Houston, TX 77058, USA}

\author[0000-0002-6615-4040]{Luisa Fernanda-Zambrano}
\affiliation{Arecibo Observatory, University of Central Florida, Arecibo, PR 00612, USA}

\begin{abstract}
The surfaces of airless bodies such as asteroids are exposed to many phenomena that can alter their physical properties. Bennu, the target of the OSIRIS-REx mission, has demonstrated how complex the surface of a small body can be. In 2019 November, the potentially hazardous asteroid \jdlong{} experienced a close approach of 0.0331~au from the Earth. We present results of the physical characterization of \jdlong{} based on ground-based radar, spectroscopy, and photometric observations acquired during 2019 November. Radar polarimetry measurements from the Arecibo Observatory indicate a morphologically complex surface. The delay-Doppler images reveal a contact-binary asteroid with an estimated visible extent of $\sim$150~m. Our observations suggest that \jdlong{} is an E-type asteroid with a surface composition similar to aubrites, a class of differentiated enstatite meteorites. The dynamical properties of \jdlong{} suggest it came from the $\nu_6$ resonance with Jupiter, and spectral comparison with major E-type bodies suggest that it may have been derived from a parental body similar to the progenitor of the E-type (64) Angelina. Significantly, we find rotational spectral variation across the surface of \jdlong{} from red to blue spectral slope. Our compositional analysis suggests that the spectral slope variation could be due to the lack of iron and sulfides in one area of the surface of \jdlong{} and/or differences in grain sizes.
\end{abstract}


\section{Introduction \label{sec:intro}}

Small rocky bodies in the solar system are a type of planetary library -- they have existed since planetary formation, undergoing minor geochemical alterations and recording information of processes that occurred ever since. Comprehending the nature (i.e., physical and dynamical properties) of small objects in the solar system has been of vital importance to understand their formation and evolution \citep{Bottke2002, Demeo2014, Granvik2018, Binzel2019}.

Wide-area asteroid surveys (i.e., the Near-Earth Object Wide-field Infrared Survey Explorer (NEOWISE, \citep{Mainzer2011}), the Panoramic Survey Telescope and Rapid Response System (Pan-STARRS; \citep{Chambers}), Spacewatch \citep{Rabinowitz1991}, the Lincoln Near-Earth Asteroid Research (LINEAR; \citep{Stokes2005}), the Catalina Sky Survey (CSS; \citep{Larson1998}), and others; see \citealt{Jedicke2015} and references therein) have been one of the principal asteroidal science data providers in the past 30~years. These surveys have significantly increased the number of detected and characterized near-Earth asteroids (NEAs), with notable implications for planetary defense. Asteroid detection and characterization (i.e., optical surveys) has provided mission support by identifying spacecraft mission targets and has enabled population studies to refine our understanding of past and future interactions among a variety of different sized objects in our solar system. However, surveys are not intended to provide a detailed characterization of individual NEAs, which is commonly achieved with subsequent ground-based telescopic follow-up.

The recent spacecraft-visited potentially hazardous asteroids (PHAs) Bennu \citep{Dellaguistina2021} and Ryugu \citep{SUGIMOTO2021114591} have revealed spectrally distinct bright boulders on their surfaces. Furthermore, these spacecrafts observations raise the question of whether it is common for NEAs to possess heterogeneous surfaces. Examples of some asteroids that have shown spectral variability include (4)~Vesta \citep{Gaffey1983, Reddy700} ($D=525.4$~km; \cite{Russell2012}), (16)~Psyche \citep{Sanchez2017} ($D=226$~km; \cite{SHEPARD2017}), 3200~Phaethon \citep{LAZZARIN2019} ($D=5.5$~km; \cite{Taylor2019}), and the active asteroid (6478)~Gault \citep{Marsset_2019} ($D=2.8$~km; \cite{Devogele2021}). Radar circular polarization ratio variation was also observed across the surface of 2006~AM$_{4}$ \citep{Virkki2014a} ($D\sim0.17$~km).

Multi-wavelength rotationally-resolved ground-based observations are key for detecting, confirming, and characterizing unique surface features (i.e., longitudinal spectral variability on asteroids). Identifying asteroids with particular regolith properties would be crucial to pin down targets for future space missions and planning and to develop accurate mitigation strategies (i.e., planning an orbit deflection to a highly porous rubble-pile versus a monolith asteroid would require special considerations of the regolith nature).

Here, we present results from our parallel multi-wavelength and multi-facility observational campaign of the NEA \jdlong{} (hereafter \jd{}). \jd{} is an Apollo-like NEA, classified as a PHA by having an Earth Minimum Orbit Intersection Distance (MOID) of 0.023~au and an absolute magnitude, $\mathrm{H}$, of 20.62 (asteroids with MOID $<$ 0.05~au and $\mathrm{H}$ $<$ 22 are classified as PHAs)\footnote{\url{https://cneos.jpl.nasa.gov/about/neo_groups.html}}. Previous observations of \jd{} found a rotation period of $5.2116\pm$0.0006~hours \citep{Warner2020} and an L-type taxonomic classification \citep{Perna2018} based on visible spectra.


\section{Observations and Data Reduction \label{sec:observations}}

We performed near-simultaneous observations of \jd{} using the Lowell Discovery Telescope and Arecibo Observatory in November 2019. Observations from Las Cumbres Observatory during the 2019 apparition are also presented in this work. This multi-wavelength approach allows us to use complementary techniques to constrain a range of physical properties for this object including its apparent shape, size, scattering properties, composition, taxonomic classification, density, and geometric albedo. Observational circumstances for Arecibo data are provided in Table \ref{tab:arecibo} and for Lowell Discovery Telescope and Las Cumbres Observatory in Table \ref{tab:ldtlco}.

\subsection{Arecibo Radar Observations}

\begin{figure*}[ht]
\includegraphics[width=\textwidth]{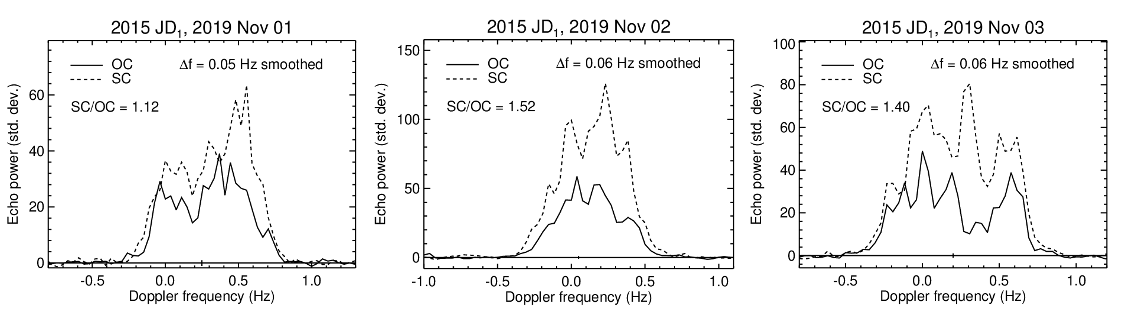}
\caption{Doppler-only echo power spectra of \jd{} from 2019 November 1 through 3 using the Arecibo S-band planetary radar system. Echo strength in both the opposite-circular (OC; solid line) and same-circular (SC; dashed line) polarization as transmitted is shown. The echo power on the Y-axis is measured in standard deviations of the background noise. The effective frequency resolution ($\Delta$f) is smoothed to 0.05/0.06~Hz.. Over the three days of observations diffuse scattering dominates the reflection.\label{fig:cw}}
\end{figure*}

\begin{deluxetable*}{cccccccc}[ht]
\tablenum{1}
\tablecaption{Radar observations of \jd{} with the Arecibo planetary radar system. Column 1 shows the UT start date and time of the observation, column 2 shows the types of observations performed. The time ($\mu$s) translates the spatial resolution of the phase modulated setup (such time does not apply to the CW setup). Column 3 shows the total observational time ($\Delta$t, includes the time duration of both CW and ranging observations), column 4 shows the average transmitted power in kilowatts, column 5 shows the distance in astronomical units, column 6 shows the observed CW Doppler frequency bandwidth in Hertz, column 7 shows the maximum visible range in meters, and column 8 shows the number of completed transmit-receive cycles for CW setup and ranging (the number of ranging cycles correspond to the total number, in case of multiple phase modulations). \label{tab:arecibo}}
\tablewidth{0pt}
\tablehead{
\colhead{\textit{Arecibo}}\\
\colhead{Start Date} & \colhead{Observation} & \colhead{Duration} & \colhead{Power} & \colhead{$\Delta$} &
\multicolumn1c{B} & \colhead{Visible Range} & \colhead{Runs}\\
\colhead{and Time (UT)} & \colhead{Types} & \colhead{$\Delta$t [min]} & \colhead{[kW]} & \colhead{[au]} &
\multicolumn1c{[Hz]} & \colhead{Extent [m]} & \colhead{[CW, Ranging]}
}
\startdata
2019-11-01 22:31:55 & CW, 1$\mu$s & 90.85 & 311 & 0.035 & 0.9 & 150 & 10, 9 \\
2019-11-02 22:53:49 & CW, 1$\mu$s, 0.2 $\mu$s, 0.05$\mu$s & 147.72 & 299 & 0.034 & 1.0 & 150, 100, 150 & 20, 94\\
2019-11-03 23:31:27 & CW, 0.05$\mu$s & 24.82 & 327 & 0.033 & 0.9 & 75 & 7, 10 \\
\enddata
\end{deluxetable*}

Radar observations of \jd{} were carried out using the Arecibo Observatory S-band (2380~MHz; 12.6~cm) planetary radar system from 2019 November 1 to November 3, including when \jd{} was at its closest approach of 0.0331~au from the Earth on November 3. Both Doppler echo power spectra and delay-Doppler imaging were collected during the three days. 

A typical Arecibo radar observation of an NEA consists of transmitting a monochromatic, circularly polarized signal either unmodulated (continuous wave, CW) or modulated (ranging or images). The transmit-receive cycle consists of transmitting for one light round-trip time (RTT) to the target, then receiving the echo for an equal time. The transmitted power was $\sim$300-400 kW during the observations. For a detailed description of radar procedure see \citet{Ostro1993}. 

The received echo was recorded in the same-circular (SC) and opposite-circular (OC) polarizations. For an ideally smooth mirror-like surface, the reflected signal is expected to be returned in OC only. The returned depolarized signal indicates the presence of both quasi-specular and diffusive scattering properties from the surface. Hence, the ratio of the SC to the OC polarization, so-called circular polarization ratio (CPR), was computed to understand the degree of possible surface undulation and composition. 

A total of 37 transmit-receive cycles on the CW setup were acquired during the scheduled radar observations (see Figure \ref{fig:cw} for visualization of the weighted sum CW spectra per day). Delay-Doppler imaging was performed at 7.5-meter range resolution during the three days of observations. To obtain the radar images, we converted the echo power into z-score normalized images (see Figure \ref{fig:delay-doppler}). For each observation, we performed CW scan-by-scan verification and selected those with robust signal-to-noise ratio. Table \ref{tab:arecibo} provides a detailed summary of the radar observations. A a detailed description of the radar observations of \jd{} (i.e., systematic issues and observation methods) will be provided in Virkki et al. (in revision).

\subsection{Lowell Discovery Telescope: near-infrared spectroscopy}

\begin{figure}[ht]
    \centering
    \includegraphics[width=0.47\textwidth]{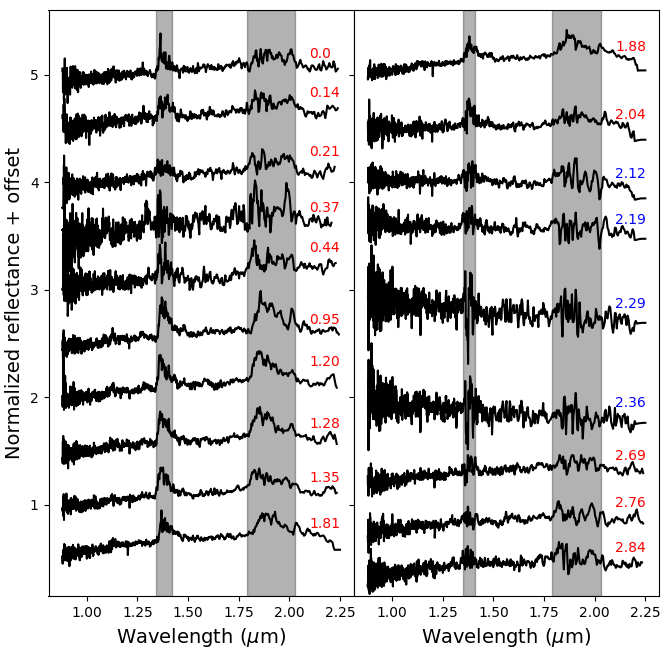}
    \caption{Rotational NIR spectra of \jd{} obtained from LDT at Happy Jack, AZ. The spectra have been organized from the beginning (first spectrum obtained at 01:39:32 UT on 11/01/2019) to the end of the observation from the top to the bottom and from the left to the right. Each spectrum has been normalized to unity at 1 $\mu$m plus an offset added in the reflectance for clarity. Grey shaded areas indicate wavelengths of telluric absorption. The numbers at the end of each spectra (around 2.2 $\mu$m) indicate the time in hours at which the spectra was obtained after the first scan. Red numbers refer to those spectra that appear red (positive slope) while blue ones refer to those whose spectra look blue (negative slope). The raw and calibration data for these rotational NIR spectra are available as data behind the figure.}
    \label{fig:dataset}
\end{figure}

Near-infrared (NIR) observations of \jd{} were conducted at the 4.3-meter Lowell Discovery Telescope (LDT) in Happy Jack, AZ, on 2019 November 1, while the asteroid was 0.037 au (14.7 lunar distances) from Earth and with a brightness of $V=15.9$ (Table \ref{tab:ldtlco}). We used the Near-Infrared High Throughput Spectrograph (NIHTS) at LDT. NIHTS is a low resolution NIR spectrograph that covers the wavelength range from 0.86 to 2.4 microns in a single order \citep{Dunham2018, Gustafsson2021}. The LDT instrument cube is equipped with a dichroic fold mirror that transmits the visible wavelengths and reflects the NIR, allowing for simultaneous visible imaging with the Large Monolith Imager (LMI) and NIR spectroscopy with NIHTS.

\begin{deluxetable*}{cccccc}[ht!]
\tablenum{2}
\tablecaption{Spectroscopy and photometry observations of \jd{}.  \label{tab:ldtlco}}
\tablehead{
\colhead{Date (UTC)} & \colhead{Time of Observations} & \colhead{Telescope} & \colhead{Phase Angle ($^\circ$)} & \colhead{V~Magnitude}
}
\startdata
2019 November 1 & 01:39:32 & LDT & 68 & 15.9 \\
2019 November 7 & 10:24:26 & LCO & 40 & 15.3 \\
2019 November 21 & 15:04:33 & LCO & 50 & 18.1 \\
\enddata
\end{deluxetable*}

The NIR spectra of \jd{} were obtained with NIHTS in the low-resolution ($R\sim50$) $4^{\prime \prime}\times12^{\prime \prime}$ slit, oriented North to South. At the beginning of the observation, we performed dome flats and arcs lamp calibrations. Dome flats were collected with dome lamps of 4700~K and 12~V inside the dome. We used the Xenon arc lamp mounted inside the instrument to generate spectral calibrations lines (see \citet{Gustafsson2021} for more details on NIHTS observing procedure). We tracked \jd{} at its sky motion rate near the meridian while the airmass ranged from 1.01 to 1.04 to minimize atmospheric dispersion. The data were acquired by nodding the object with 5$''$ offsets along the spatial direction of the slit between the A and B positions in an ABBA pattern. We used a $120$~s exposure time to optimize both signal and wavelength coverage. The ABBA nod pattern with $120$~s exposures was repeated for 2.8 h of observation resulting in a total of nineteen spectra shown in Figure \ref{fig:dataset}. The conditions during the observation remained relatively clear with some wind and thin clouds (seeing ranged from 1\farcs4 to 1\farcs7). The precipitable water value was approximately $1.1-1.8$~mm \footnote{\url{http://weather.uwyo.edu/upperair/sounding.html}}. To correct for telluric absorption and to obtain relative reflectance, the solar analog star SA113-276 was observed at similar airmass to the asteroid. The standard data reduction procedures were performed using the IDL \textit{Spextool} package \citep{Cushing} which has been adapted for NIHTS. Figure \ref{fig:dataset} shows the whole spectral data set covering $\sim$56\% of the \jd{} period ($5.2116\pm$0.0006~hours; \citet{Warner2020}).

\subsection{Lowell Discovery Telescope: photometry}

\begin{figure}[b!]
    \centering
    \includegraphics[width=0.47\textwidth]{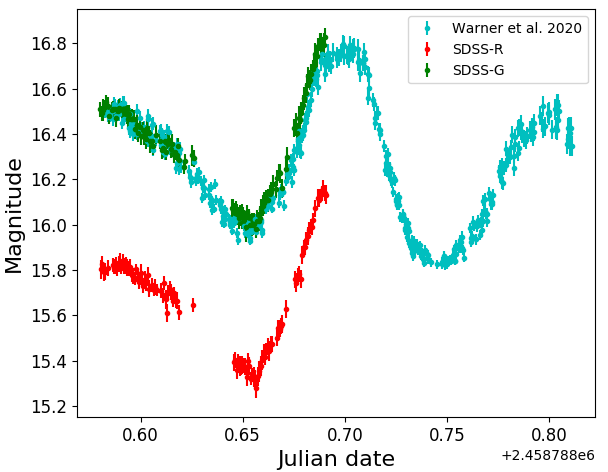}
    \caption{SDSS $g$- (green dots) and $r$-band (red dots) filters lightcurves of \jd{} as a function of time since the beginning of the observation obtained using the Large Monolithic Imager at LDT. These single-peaked lightcurves shows half of \jd{}'s bimodal lightcurve shown in cyan dots from \citet{Warner2020}. The gap in the LDT data between 1.3 and 1.7~hours correspond to the observation of the standard star. The g- and r- lightcurves are available as data behind the figure; the raw photometry data are made available on FigShare: doi:10.6084/m9.figshare.20137265.v1.}
    \label{fig:lmi_gr}
\end{figure}

We obtained simultaneous visible photometry of \jd{} with LMI through the dichroic fold mirror \citep{Massey2013}. Observations were conducted in Sloan Digital Sky Survey (SDSS) $g$- and $r$-band filters. The full LMI field of view captures 12\farcm5 $\times$ 12\farcm5 with a $6144 \times 6160$ pixel CCD camera with an unbinned pixel scale of 0\farcs12/pixel. The transmitted optical beam through the dichroic fold mirror creates a roughly oval-shaped unvignetted field of view of 6\farcm5 $\times$ 4\farcm0. Within this field, the throughput loss on $g$ and $r$ filters is $\sim$10\% \citep{Gustafsson2021}. Images were obtained in 2$\times$2 binning mode (0\farcs24/pixel) with exposure times ranging between 4~s and 5~s. Flat-field images were taken for $g$ and $r$ filters during astronomical twilight on 2019 November 1. All data were bias subtracted and flat-field corrected for their corresponding filters. 

The reduced LMI imaging data were analyzed using the automated \textit{Photometry Pipeline} (PP) \citep{Mommert}. PP first utilizes Source Extractor \citep{1996A&AS..117..393B} to identify objects within each field, then SCAMP \citep{Bertin} is called to match all sources with the selected photometry catalog. The astrometry was performed using data from the Gaia (DR1) mission \citep{Gaia2016}. Pan-STARRS \citep{Chambers} and SDSS \citep{Ahn} catalogs were individually used to perform the photometry. Astrometry and photometry catalog queries were executed via PP using the Vizier Catalog Service \citep{Ochsenbein2000Vizier}. We performed manual and visual identification of the target for the whole set of images to discard images where the target was passing over or nearby a star or cosmic ray and to ensure correct target identification during the automated process. We manually verified a random sampling of results using Aperture Photometry Tool \citep{Laher2012}. We performed another PP calibration using only stars with solar-like colors to compare with the automated photometric analysis and to obtain a more reliable photometric reduction. Ultimately, 115 and 105 data points were obtained for the $g$- and $r$-band filters, respectively. We interpolated the lightcurves in Figure \ref{fig:lmi_gr} using the \textit{interp} routine of the \textit{numpy Python} package to obtain the respective $g$ and $r$ magnitudes at equal times.

\subsection{Las Cumbres Observatory: visible spectroscopy \label{sec:lco}}

\begin{figure}[h!]
    \centering
    \includegraphics[width=0.47\textwidth]{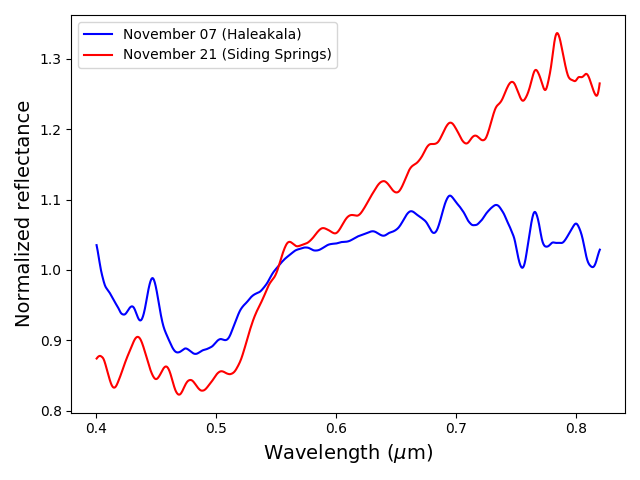}
    \caption{Visible spectra obtained at Haleakala Observatory, Kula, HI and Siding Springs, Australia from Las Cumbres Observatory. Blue and red lines are the visible spectra obtained on 2019 November 7 and November 21, respectively. Both spectra have been normalized to unity at 0.55 $\mu$m. The raw and calibration data for these visible spectra are available as data behind the figure.}
    \label{fig:lco}
\end{figure}

In this paper, we also present visible spectroscopic observations obtained by Las Cumbres Observatory (LCO) using the FLOYDS\footnote{\url{https://lco.global/observatory/instruments/floyds/}} spectrographs located on the 2.0~m telescopes in Haleakala, Hawaii, and Siding Springs, Australia, on 2019 November 7 and November 21 at UTC 10:24:26.7 and 15:04:33.1, respectively. The data shown in Figure \ref{fig:lco} are the smoothed, reduced spectra which were scheduled, observed, and reduced via a fully remote and automated process using the NEOExchange observation portal \citep{Lister2021} and the LCO automated pipeline\footnote{\url{https://lco.global/documentation/data/floyds-pipeline/}}. Exposure times of 1800~s and 3600~s were used while airmass ranged from 1.4 to 1.6 and 1.1 to 1.3 for the November 7 and November 21 observations. The solar analogs HD60298 and SA93 were observed near the target for November 7 and November 21 observations. On November 21, \jd{} Vmag was $\sim$18.1, resulting in a poorer-quality observation. 

\begin{figure}[h!]
    \centering
    \includegraphics[width=0.47\textwidth]{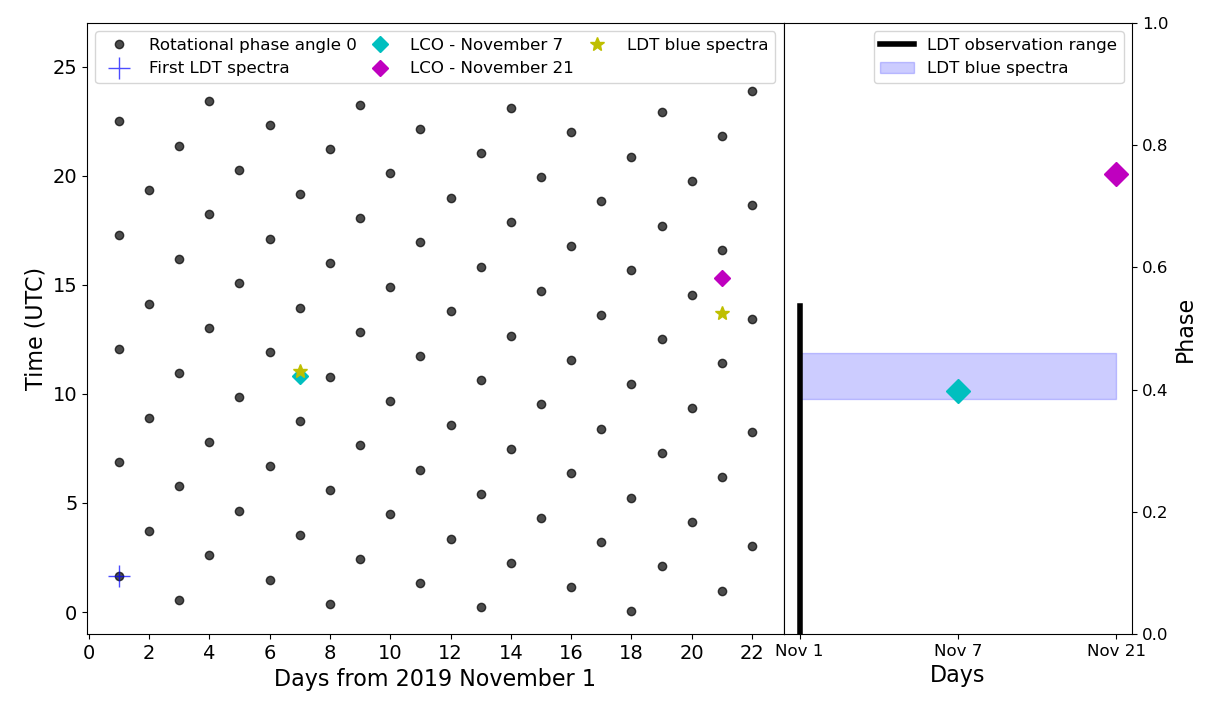}
    \caption{Left: track of \jd{}'s rotational phase angle 0$^\circ$, defined as the orientation of the first LDT spectra obtained at 01:39:32.37 UTC on November 1, as a function of days (grey filled circles). Yellow stars indicate the time when observing the blue spectra at LDT. The cyan and magenta diamonds illustrate the LCO observations. Right: phase range of each observation per day of the observation. The blue shaded area illustrates the phase range of blue NIR spectra at LDT. The black line shows the range of the LDT observation, while diamonds represent each LCO observation in the corresponding phase. The measurement uncertainties are shown in the text.}
    \label{fig:phasing}
\end{figure}

We phased the LCO observations to our LDT observation to check if we were likely observing the same object geometry. In Figure \ref{fig:phasing} we present the tracking of the zero (0$^\circ$) rotational phase angle ($\Theta_{0}$, beginning at UTC 01:39:32.37 on November 1) or the orientation that we were observing on the first scan obtained at LDT-NIHTS by adding \jd{}'s rotation period ($5.2116\pm$0.0006~hours; \citet{Warner2020}). Because LCO observations were performed several days after our LDT observation, we also considered the relative offset induced by sky motion when phasing these observations. On November 7 and November 21, we find that LCO observations were $28^\circ$ and $18^\circ$, respectively, different in phase angle from where we were looking with LDT on November 1 (68$^\circ$). The difference in phase angles imposes an offset of $+$25 and $+$15~minutes to the derived $\Theta_{0}$ for November 7 and November 21, assuming a prograde rotation. We find that the LCO spectrum from November 7 (see the cyan diamonds in Figure \ref{fig:phasing}) corresponds to the orientation we were observing for \jd{} at the LDT around 2.36$\pm$0.02~hours after $\Theta_{0}$ (see the yellow star in the left panel or the blue shaded area in the right panel). The observation from November 21 (see the magenta diamonds) covered an orientation not observed by the LDT observations, specifically 4.86$\pm$0.06~hours after $\Theta_{0}$ or 2.02~hours after the last spectrum presented in Figure \ref{fig:dataset}. In summary, the bluer LCO visible spectrum from November 7 coincides with the orientation we were observing for the blue sloped NIR spectra at LDT on November 1, while the steeper red LCO spectrum corresponds to an orientation that we did not cover at LDT (see the right panel of Figure \ref{fig:phasing} for better data visualization). Furthermore, both LCO and LDT observations support spectral variability in a similar near-surface region on \jd{}.

\section{Results \label{sec:results}}

\subsection{Constraining Composition and Physical Properties}

\begin{figure*}[t!]
    \centering
    \includegraphics[width=\textwidth]{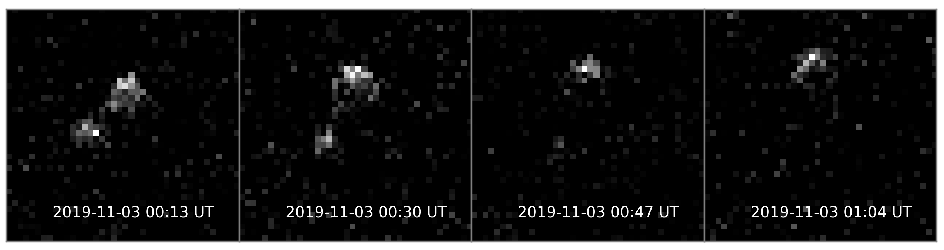}
    \caption{Compilation of Arecibo delay-Doppler images of \jd{} from 2019 November 3 showing roughly a sixth of its full rotation. On each frame, the resolution is $7.5$~m/px (vertical) $\times$ 0.0745~Hz/px (horizontal). Each radar image is the weighted sum of fifteen consecutive transmit/receive cycles. Time increases from left to right and is indicated at the bottom of each panel.}
    \label{fig:delay-doppler}
\end{figure*}

\begin{figure}[b!]
    \centering
    \includegraphics[width=0.47\textwidth]{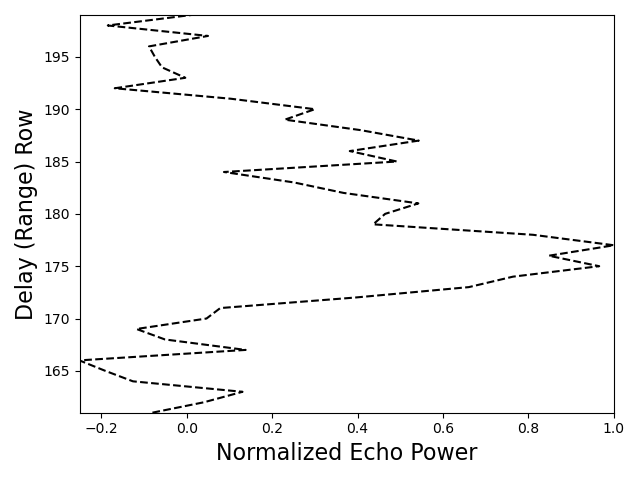}
    \caption{Distribution of echo power as a function of normalized echo depth, made by collapsing the horizontal dimension of the range-Doppler image. The strong signal in at least 20 rows at 7.5 m/pixel suggest $\sim$150~m along the long axis.}
    \label{fig:extent}
\end{figure}

Arecibo delay-Doppler images reveal a contact-binary asteroid with an elongated projection, as expected from the 0.9~mag lightcurve amplitude \citep{Warner2020}. The bi-lobed structure presented in Figure \ref{fig:delay-doppler} shows a body and a head of approximately 100 and 50~m, respectively, or around 150~m in visible extent (see Figure \ref{fig:extent}). Radar images and CW experiments show no evidence of separate binary structure, which usually appears as a delta peak in the Doppler-only echo power spectra \citep{NAIDU2020}.  

The disk-integrated CPR, the quotient of the diffuse and the quasi-specular integrated echo power (SC/OC) is 1.35 $\pm$ 0.54 on average and 1.12 $\pm$ 0.29, 1.52 $\pm$ 0.40, and 1.4 $\pm$ 0.22 on November 1, November 2, and November 3, respectively (see Figure \ref{fig:cw} for data visualization). \jd{} has one of the highest and consistent disk-integrated CPR values ever reported among other radar NEAs \citep{Benner2008, Aponte2020}. We should point out that the system temperature of the Arecibo S-band radar system suffered of some issues due to the low-noise amplifier (LNA) degradation after April 2019. The LNA issue could have produced lower measurements of OC cross-sections, thus yielding additional uncertainty on the CPR values. We attempted to address this issue in the data reduction. It seems statistically unlikely that \jd{}'s high CPR measurements were affected by this issue because of the consistency over the three days of radar observations. More discussion about the LNA issue and how it may have affected Arecibo's S-band radar observations is soon to be published by Virkki et al. (in revision).

The measured total radar cross-section $\sigma_{T}$, given by the sum of the geometric OC and SC cross-sections, of \jd{} on November 2 and November 3 is 6356$\pm$146~m$^{2}$ and 11911$\pm$158~m$^{2}$, respectively, with a systematic calibration uncertainty of 25\%. In other words, for example, if \jd{} were a perfectly isotropically reflecting metal sphere of 6356m$^{2}$ geometric cross-section it would return the same amount of power if observed under similar circumstances. Assuming \jd{} is a sphere of 0.15 km its upper limit OC ($\hat{\sigma}_{OC}$) and SC ($\hat{\sigma}_{SC}$) radar albedo (the ratio of the radar cross-section, of a given polarization, by the derived effective projected area) would be 0.22$\pm$0.05 and 0.31$\pm$0.08, respectively, or a total radar albedo $\hat{\sigma}_{T}$ = 0.53$\pm$0.10. Additional uncertainty in radar albedo could arise from deviation in the projected area (i.e., highly elongated asteroids), which can add an offset of 25\% to the radar albedo computed above.

\subsection{Sub-radar point}

The spin-axis of a celestial body can be constrained from radar by measuring the change in the sub-radar latitude point. As the asteroid rotates, one limb approaches the line-of-sight while the other rotates away such that the transmitted monochromatic signal is broadened in Doppler. For a spherical with rotation rate $P$, the broadened received signal in an echo bandwidth $B$ is given by

\begin{equation}
 B(\phi) = \frac{4\pi D(\phi)}{\lambda P} cos(\delta)
 \label{eq:bandwidth}
\end{equation}

\noindent where $D$($\phi$) is the breadth of the asteroid at rotation phase $\phi$, $\lambda$ is the wavelength of the transmitted signal (0.126 m), and $\delta$ is the sub-radar latitude on the asteroid as viewed by the observer. Although \jd{} is not a spherical object, this expression holds for any irregular (non-spherical) object. We modeled \jd{}'s Doppler dispersion bandwidth (shown in Figure \ref{fig:cw}) using equation \ref{eq:bandwidth}. 

In Eq. \ref{eq:bandwidth}, the sub-radar latitude variability is limited by the change of Doppler dispersion when the rotation period and the sphere diameter is well-established. Figure \ref{fig:cw} shows small Doppler dispersion differences at each observation, which will induce negligible sub-radar latitude changes. We found that an effective diameter of 150~m with a rotation period of 5.2116~h yield a Doppler dispersion of $\sim$0.9 Hz if \jd{} was observed at or near its equator (around $\pm$70 degrees from the pole). The largest observed bandwidth of $\sim$1.2 Hz on November 3 requires an effective diameter of around 190~m also suggesting a near-equator sub-radar latitude point. Furthermore, our solutions support the rotation period derived by \citet{Warner2020}. In addition, Figure \ref{fig:delay-doppler} shows the smaller lobe disappearing as the body rotates, which is evidence of a sub-radar latitude near the equator (i.e., if the sub-radar point were far off the equator, both lobes would be visible the whole time).

\subsection{Spectroscopic variation and photometric color \label{sec:variability}}  

\begin{figure}[ht!]
 \begin{center}
  \includegraphics[width=0.47\textwidth]{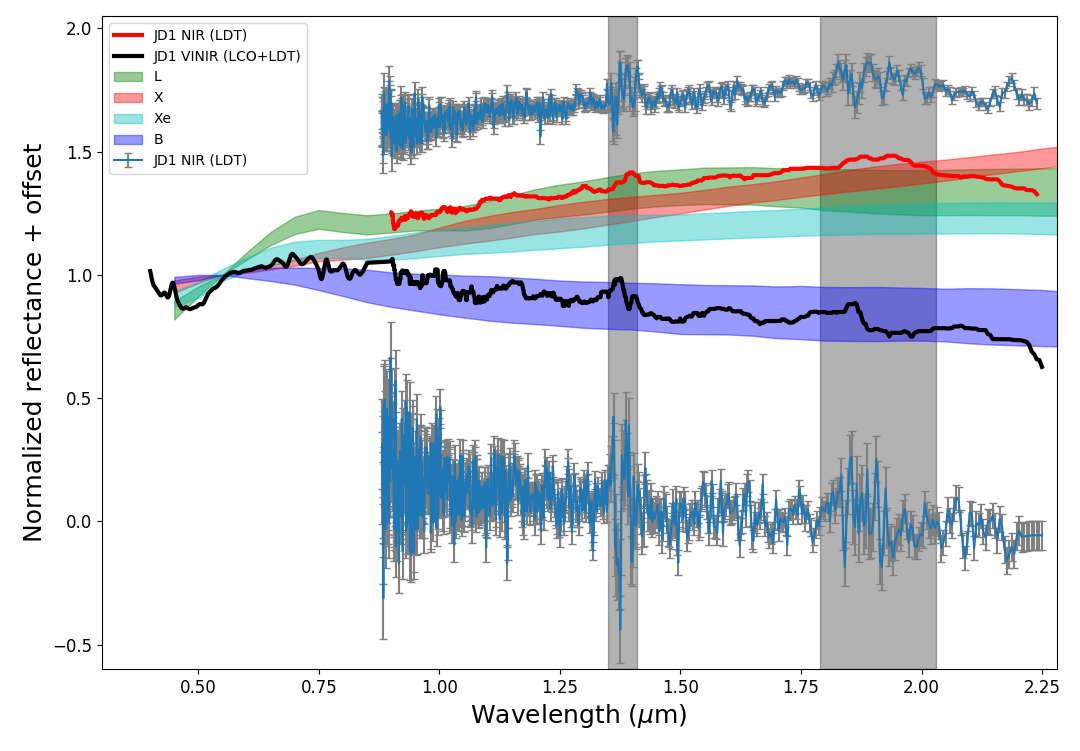}
  \caption{Visible and NIR spectra of \jd{} obtained at LCO and LDT. Top spectrum (blue solid line): \jd{} positive-sloped (red) featureless non-smoothed spectra. Bottom spectrum (blue solid line): negative-sloped (blue) featureless non-smoothed spectra. The red, cyan, green, and blue shaded areas are the $\pm$ one-sigma spectra of X-, Xe-, L-, and B-type asteroids, respectively, from Bus-DeMeo taxonomy. Bus-DeMeo and \jd{} visible and NIR spectra have been normalized at 0.55 $\mu$m. The grey shaded areas indicate residual telluric features from the data reduction. Both NIR spectra from LDT (blue line spectra) have been normalized to unity at 1 $\mu$m plus an offset in reflectance for clarity. \label{fig:XB}}
 \end{center}
\end{figure}

In Figure \ref{fig:XB}, we present two of nineteen featureless NIR spectra of \jd{}: a positive-slope (top) and a negative-slope (bottom) spectrum. These spectra were selected to illustrate the differences between their slopes. The negative-sloped (blue) and positive-sloped (red) NIR spectra correspond to the spectra obtained at 2.29 and 2.69~hours at LDT with NIHTS shown in Figure \ref{fig:dataset}. In the black solid lines we show the visible and NIR spectra of \jd{}. The NIR spectra were smoothed using a median filter from \textit{Scipy} Python package and offset to match the normalized visible reflectance from LCO. The non-smoothed red and blue NIR spectra (blue line) are plotted above and below to their respective smoothed spectra. The most likely Bus-Demeo spectral classifications are plotted in Figure \ref{fig:XB} as red, cyan, green, and blue shaded area for X-, Xe-, L-, and B-type, respectively, and suggest spectroscopic similarities between these taxonomies: X-complex and B-type classification for the red and blue spectrum of \jd{}, respectively. 

\citet{Perna2018} classified \jd{} as an L-type asteroid. L-type spectra are characterized by the positively sloped visible spectra with a gentle concave curvature around 1.5 $\mu$m. The red visible and NIR spectra of \jd{} appear like an L-type but also are not too different from an Xe-type. Distinguishing between these two taxonomies is often very complicated because of the spectral similarities. \jd{}'s NIR spectra lack concave and convex curvatures around 1.5 and 2.0 $\mu$m, respectively (see Figure \ref{fig:dataset}), which are characteristic features of L-types \citep{Devogele2018}. Additionally, the absorption band feature of \jd{} around 0.5 $\mu$m (see Figure \ref{fig:lco}) is typical of sulfide minerals such as oldhamite, seen on E-types \citep{Clark2004a}. This spectral analysis combined with our radar measurements suggest that \jd{} is likely an X-complex/E-type. Although our observations suggest an E-type taxonomy, the red spectra of \jd{} is closely resembled to L-types. It is critical to mention that L-types are extremely rare primitive asteroids which constitute a very small fraction of objects in the main belt of asteroids \citep{Devogele2018}.

Laboratory and telescopic studies have proved NIR spectral variability due to differences in phase angles, atmospheric dispersion, grain sizes, surface refreshing, observational circumstances (i.e., instrumental failure, target out of slit, or stellar interference), space weathering, and/or mineralogical composition. 

\begin{figure}[hb!]
\includegraphics[width=0.47\textwidth]{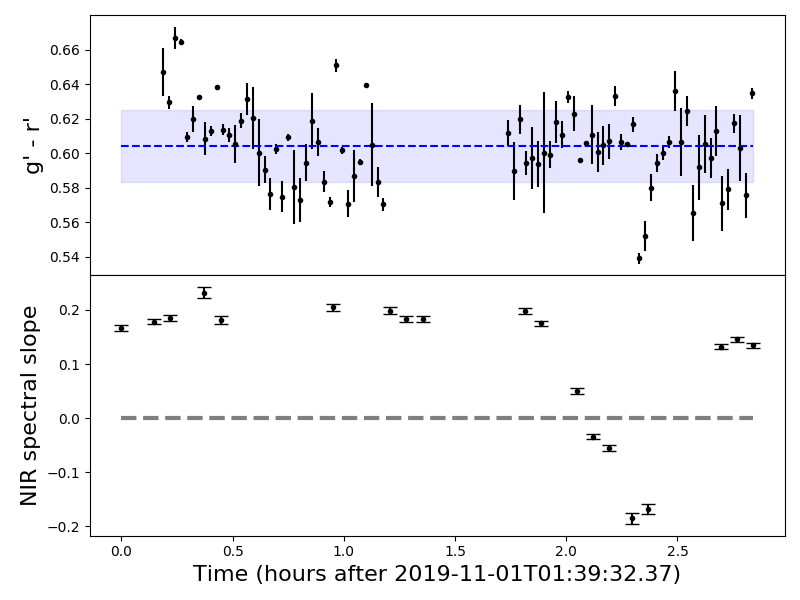}
\caption{Top panel: Sloan g$^{\prime}$ - r$^{\prime}$ color variation as a function of time since the beginning of the LDT observation. Horizontal dashed blue line and blue shaded area indicates the average g$^{\prime}$ - r$^{\prime}$ color and the standard deviation, respectively. Bottom panel: NIR continuum-spectral slope variation as a function of time since the beginning of the observation. Slope variation can be noticed after $\sim$1.9~hours.\label{fig:variation}}
\end{figure}

The NIR spectral slope variability observed in \jd{} (bottom panel, Figure \ref{fig:variation}) is transitional: a consistent positive-slope observed for about 2~hours after the first measurement, changing to a blue negative-slope for $\sim$0.5 hours and eventually transitioning back to red positive-sloped spectra. The variability suggests that any of the previous factors could be affecting our spectroscopic result. However, we can conservatively discard some of them.

Spectroscopic studies (i.e., \citet{Reddy2012, Nathues2010, Sanchez2012}) have shown the increase/decrease of spectral slope and band depth when the phase angle g, increase between 0$^\circ$ $<$ $g$ $<$ 120$^\circ$. \jd{}'s phase angle decreased from 68$^\circ$ to 67$^\circ$ from the beginning to the end of the observation at the LDT. The change in phase angle is about 0.5$^\circ$, too small to be responsible for the observed spectral variability. Thus, we can confidently discard the phase angle factor. 

\begin{figure}[h!]
    \includegraphics[width=0.47\textwidth]{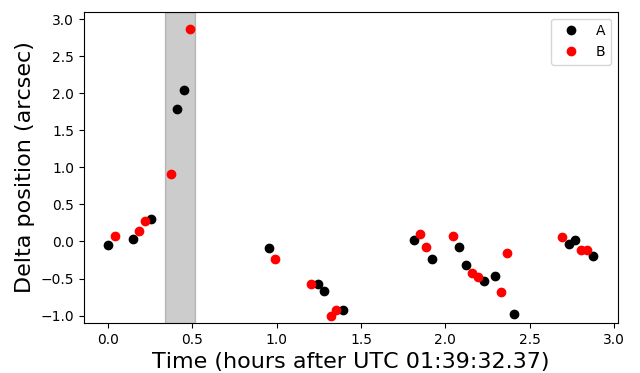}
    \caption{Pixel shift of \jd{} as a function of time (starting at 01:39:32 UTC). Black and red circles are the delta central aperture position of each spectrum at the $A$ and $B$ slit position, respectively. The grey shaded area indicates the timing of the largest slit drift (spectral variability occurs after $\sim$1.9~hours). \label{fig:shift}}
\end{figure}

The observed spectral variability (bottom panel, Figure \ref{fig:variation}), does not suggest a phase-angle-like variability, but opens the possibility of an observational failure, i.e., the asteroid moving out of the slit. If the slit was not consistently aligned in the same way for every exposure either in position A or B, then, in theory, there could be a shift in intensity that is wavelength dependent. If the slit was fixed, as during our observation, any fraction of light lost due to a finite slit is the same for every exposure at each wavelength. We used \textit{Spextool} to measure the central aperture slit position as a function of time. Figure \ref{fig:shift} shows \jd{}'s aperture shift in the $4^{\prime \prime}$ slit as a function of time. The timing of the largest drift does not line up with the change in NIR spectral slope we found in \jd{}. At the time of observing spectral variability, we found a drift of $<1^{\prime \prime}$ ($\sim$9~pixels) between the asteroid and the slit. Thus, the possibility that the asteroid was moving out of the $4^{\prime \prime}$ slit during the time of spectral slope change can be rejected.

Another factor that we considered is the atmospheric dispersion. However, this could be easily discarded, given that \jd{} was observed at relatively low airmasses between 1.01 to 1.04. We verified the standard star slopes and found similar normalized spectral slopes with no difference larger than 2\% in continuum. We found consistency in our data reduction (i.e., no correlation between standard star slopes and \jd{}'s variability). We performed an inspection of our spectral data and found no stellar interference contaminating \jd{}'s flux within the $4^{\prime \prime}$ slit in our dataset. 

With the provided overview, it does not seem feasible that our data was affected by observational artifacts. We explore other possibilities that could be causing such spectral variabilities in the next Sections. However, such possibilities support the position that the observed variability truly reflects surface properties of \jd{}. 

\begin{figure*}[ht!]
    \centering
    \includegraphics[width=\textwidth]{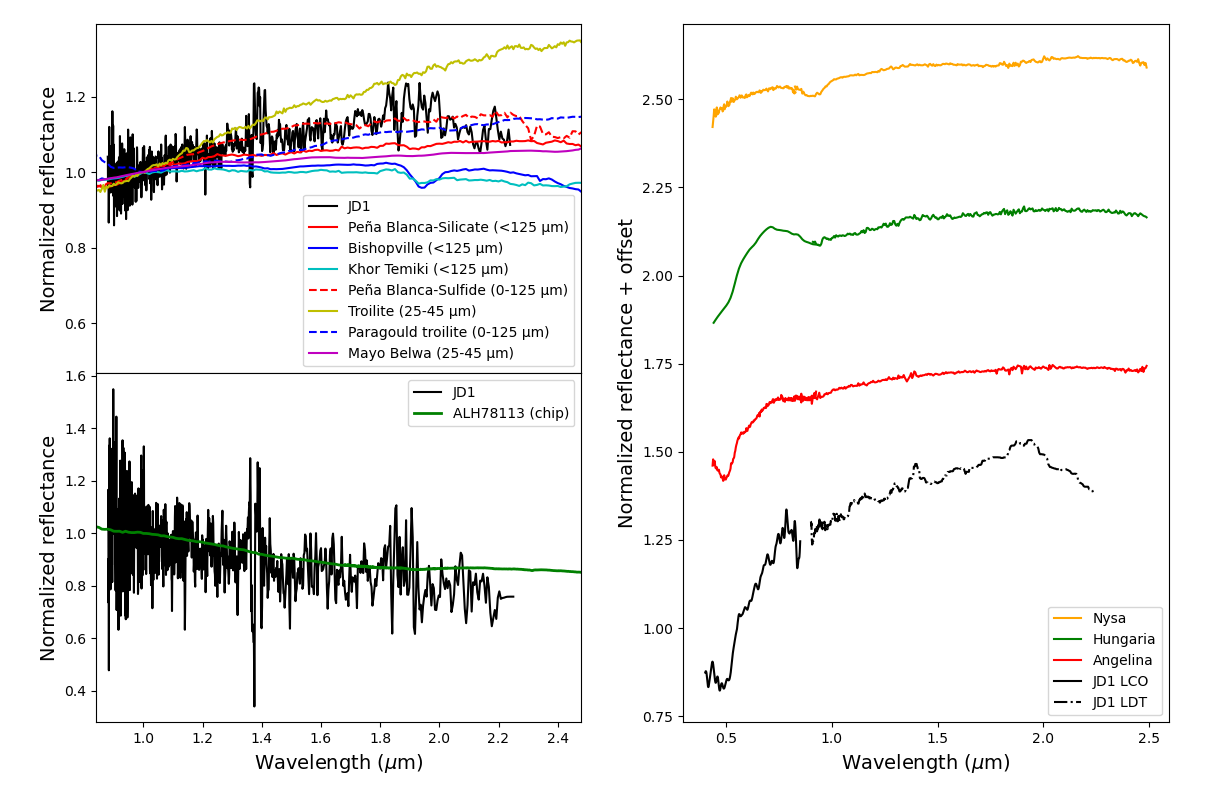}
    \caption{Plot of NIR spectra of \jd{} (continuous black line) and selected RELAB meteorites, minerals, and E-type asteroids. Left top panel: reflectance of aubrites Mayo Belwa (RELAB sample code s1tb46), Khor Temiki (RELAB sample code c1tb48), Bishopville (RELAB sample code c1tb47), Peña Blanca silicate and sulfide regions (RELAB samples codes c1tb45 and c1tb55, respectively) and minerals troilite and Paragould troilite (RELAB sample code cdmb06 and latb50, respectively). Left bottom panel: NIR blue reflectance of \jd{} and aubrite ALH78113 (RELAB code bkr1ar001). Right panel: \jd{} visible and NIR spectra and E-type asteroids (434) Hungaria, (44) Nysa, and (64) Angelina obtained from \citet{Binzel2001}, \citet{Clark2004a}, and \citet{Bus2002}, respectively, plus an offset in the reflectance for clarity. The NIR reflectance of \jd{} was smoothed using a median filter from \textit{Scipy} Python package and offset to match the visible reflectance. All the spectra in the left and right panel were normalized at 1 $\mu$m and at 0.55 $\mu$m, respectively. \label{fig:M&A}}
\end{figure*}

The average $g^{\prime}$ - $r^{\prime}$ color of \jd{} is 0.60 $\pm$ 0.03, shown in top panel of Figure \ref{fig:variation} in the blue dashed line. This color, within errors, falls in the range of $g^{\prime}$ - $r^{\prime}$ colors (0.4-0.6) for X-types \citep{Ivezic}. It is well known that asteroid colors reflect properties of their composition. We also look for visible color variation in order to validate any spectral variation seen in the NIR. The color variation, shown in black dots at the top panel of Figure \ref{fig:variation}, support color blueing from 0-0.7~hours and around 2.35~hours. While there is a visible color change in \jd{} around 2.35~hours, it is statistically of low significance to support the NIR spectral slope variability, also considering the noisy rotational color distribution. The largest visible color drop occurs around the bluer NIR spectra, while no color decrease is observed at 1.9 hours when the NIR slope changes. This could be due to a non-simultaneous spectroscopic variability between the visible and NIR region or to the noise in the photometric color distribution. 

While the completeness and reliability of \jd{} NIR spectral variation are supported by a single observation night of $<$60\% of \jd{}'s period, and thus not confirmed, independent observations from LCO also suggest visible spectral variability. We demonstrated in Section \ref{sec:lco} that the geometry observed with the blue NIR spectra from LDT coincide with that of the blue visible LCO spectrum. Therefore, we interpret the NIR slope variability as a true reflection of particular surface properties of \jd{} and not as artifacts in the observations. Still, further observational validation is needed.

\section{Analysis \label{sec:compositional analysis}}

\subsection{Visible albedo \label{sec:vis_albedo}}
We investigated the propagation of visible albedo solutions considering two different diameters and three absolute magnitudes (H) for each diameter (see below the details of the adopted methodology). 

It is known that H mag is the main driver to derive the albedo uncertainty because of the difficulty in measuring the entire asteroids’ phase curve \citep{Masiero2021}. We used the H-G \citep{Bowell1989} system to calculate the absolute magnitude $H_{f}$ at a given photometric filter f. We used our measured visible magnitudes at the SDSS $g$ and $r$ band filters, V$_{g}=$16.25 and V$_{r}=$15.65, a G slope value of 0.15, and phase angle $\alpha$=68$^{\circ}$. We obtained g and r bands absolute magnitudes H$_{g}=21.03$ and H$_{r}=20.36$, respectively. In our visible albedo analysis, we also considered the Minor Planet Center absolute magnitude H$_{MPC} = 20.62$. We used the \citep{Fowler1992} empirical expression which relates the absolute magnitude H, diameter, and visible albedo of an asteroid. By using a fixed diameter of 150~m we obtained a visible albedo ranging from 0.30 to 0.56 and 0.19 to 0.35 assuming a 190~m effective diameter (see table \ref{tab:albedos} for detailed solutions). The range of possible visible albedo solutions gives an average albedo $p_{ave}$ of 0.35$\pm$0.12. The average visible albedo, within the standard deviation, falls in the range of albedo for enstatite-like asteroids ($\sim$0.4 to $\sim$0.6) \citep{Zellner1977, Zellner1985, Clark2004a, Thomas2011} and with the 0.50$^{+0.30}_{-0.24}$ derived by \citet{Trilling2016} for \jd{}.

\begin{deluxetable}{cccc}[h!]
    \tablenum{3}
    \tablecaption{Visible albedo ($p$) solutions of \jd{} for a range of effective diameter and absolute magnitudes. \label{tab:albedos}}
    \tablewidth{0pt}
    \tablehead{& \multicolumn3c{Absolute magnitudes}\\
    \colhead{} & \colhead{H$_{g}$} & \colhead{H$_{MPC}$} & \colhead{H$_{r}$} \\ 
    \colhead{} & \colhead{21.03} & \colhead{20.62} & \colhead{20.36}
    }
    \startdata
    Visible albedo & & & \\
    \textit{p($D=150~m$)} & 0.30 & 0.44 & 0.56 \\
    \textit{p$(D=190~m)$} & 0.19 & 0.28 & 0.35 \\ 
    \textit{p(average)} & \multicolumn3c{0.35 $\pm$ 0.12} \\
    \hline
    \hline
    \enddata
\end{deluxetable}

\subsection{Meteorite and asteroid connections \label{sec:M-A}}

We have presented evidence that supports a relationship between \jd{} and E-types asteroids, such as the high albedo of \jd{}, high CPR, spectral shape similarities, and near-surface bulk density. We performed a search through the Reflectance Experiment Laboratory (RELAB)\footnote{\url{http://www.planetary.brown.edu/relab/}} database to constrain the composition and trace the possible meteorite connection with \jd{}. We searched for enstatite achondrites meteorites (aubrites) and common minerals usually present on these meteorites, in which NIR spectral properties were mostly featureless \citep{Cloutis1990}.

Aubrites are well-known brecciated meteorites chemically dominated by low-iron enstatite (75-98 vol\%), with relatively small amount of plagioclase (4.3 vol\% on average), and nearly FeO-free diopside (2.2 vol\% on average) and forsterite (3.5 vol\% on average) \citep{Watters1979, Gaffey1983, Keil2010}. Bulk chemical analysis indicates that oxides such as MgO and SiO$_{2}$ are the most abundant, while CaO, Na$_{2}$O, and Al$_{2}$O$_{3}$ are less abundant. Other metallic elements (Fe and Ni) and sulfides (Ti, Mn, Cr) are also found to be less abundant. The chemical composition overview of aubrites gives an idea of what elements should be expected on E-types. In Figure \ref{fig:M&A} (left panel), we show the reflectance of well-known aubrites and common minerals on them. By visible inspection, spectral slopes similarities can be noticed between \jd{} red and blue spectrum and aubrites such as Peña Blanca and ALH78113.

We have implemented the E-type sub-classification techniques developed by \citet{Gaffey2004} and \citet{Clark2004a} to investigate the composition of \jd{}. \citet{Gaffey2004} proposed three sub classes for E-types: E[I], E[II], and E[III]. The E[I] show a slightly curved spectrum with no clear mineral absorption feature. The E[II] are distinguished by the strong absorption features near 0.5 $\mu$m and a weaker one near 0.9 $\mu$m. These absorptions are characteristic of oldhamite. The E[III] are characterized by having an almost-flat spectrum which turns to a red slope with a unique absorption feature near 0.9 $\mu$m, attributed to enstatite pyroxene with some Fe$^{2+}$. In addition, \citet{Fornasier2008} found that E[II] asteroids had the highest spectral slopes while E[III] asteroids had flat or negative visible spectral slopes. The \citet{Clark2004a} sub-classification technique is based on mineralogical compositional similarities to the E-types asteroids (44) Nysa, (64) Angelina, and (434) Hungaria. The Nysa-like are those whose compositional analysis suggests silicate mineralogy with higher iron content than enstatite and absorption features at 0.9 $\mu$m and 1.8 $\mu$m. The Angelina-like are characterized by having absorption features at 0.5 $\mu$m and 0.9 $\mu$m, attributed to silicate mineralogy plus the presence of sulfide and olivine. The Hungaria-like are asteroids whose compositional analysis revealed some olivine and a typical reflectance feature near 0.9 $\mu$m.

The \citet{Gaffey2004} classification suggests that \jd{}, having a possible absorption feature centered at 0.507 $\mu$m and lacking the 0.9 $\mu$m feature, is consistent with an E[II] classification. In Figure \ref{fig:M&A} (right panel), \jd{} red spectrum is plotted with the E-types (434) Hungaria, (64) Angelina, and (44) Nysa for comparison. \jd{} has a red spectra among the E-types with a particular spectral similarity to (64) Angelina. Using both \citet{Clark2004a} and \citet{Gaffey2004} classifications, we suggest that \jd{} corresponds to the Angelina-like group, given their spectral similarities observed in Figure \ref{fig:M&A}. The presented spectral connection implies that \jd{} mineralogy should be similar to that of aubrite meteorites plus oldhamite and troilite \citep{Clark2004a}.

\subsection{Grain sizes \label{sec:grain}}

Spacecraft-visited asteroids (\textit{i.e.}, Itokawa, Bennu, Eros, and Ryugu) have shown that the surfaces of small bodies are highly diverse, with particles ranging from dust-like to boulder-like sizes. Such particle variability could affect the deduced spectral composition by making spectra appear distinct at different phase orientations. Meteorite reflectance as a function of composition and grain-sizes would provide crucial information on whether a spectral variation is due to different grain-sizes or different mineralogical composition or a combination of both. 

\begin{figure*}[ht!]
\includegraphics[width=\textwidth]{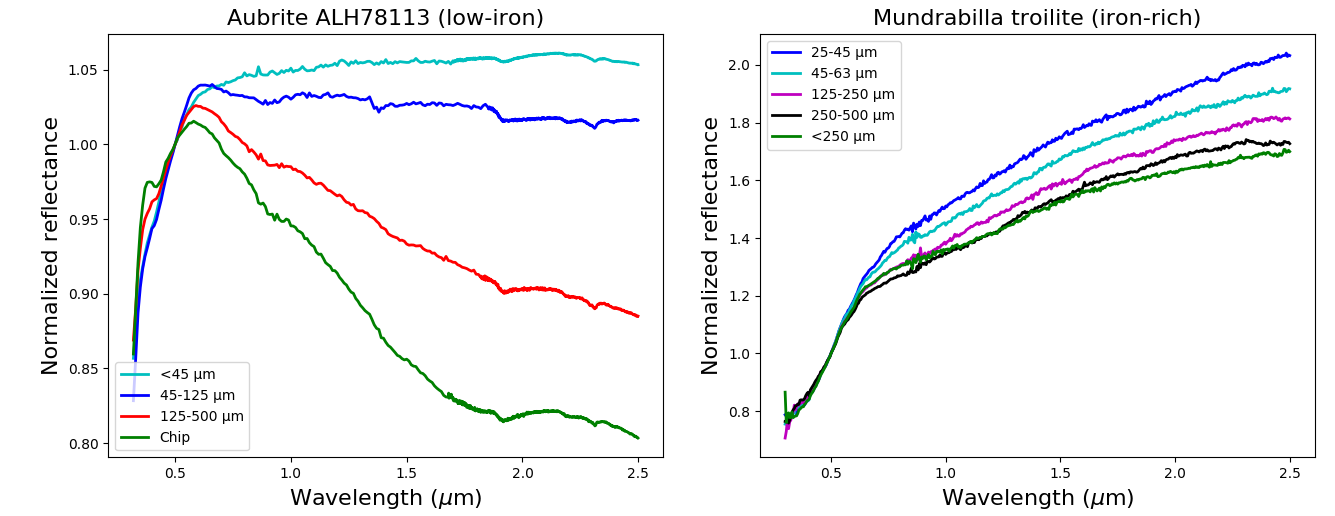}
\caption{Plot of RELAB meteorites reflectances showing the effect of grain sizes in the NIR slope. Left: reflectance of the aubrite ALH78113 at different grain sizes (low-iron sample). RELAB samples codes are bkr1ar001, bkr1ar001a, bkr1ar001b, and bkr1ar001c for ALH78113 reflectances with grains sizes in the range of ``chip", 125-500 $\mu$m, 45-125 $\mu$m, and $<$45 $\mu$m, respectively. Right: reflectance spectra of Mundrabilla meteorite at different grain sizes (iron-rich sample). RELAB samples codes are cbmb06, ccmb06, cemb06, cfmb06, and c1mb06 for Mundrabilla's reflectances with grain sizes in the range of 25-45 $\mu$m, 45-63 $\mu$m, 125-250 $\mu$m, 250-500 $\mu$m, and $<$250 $\mu$m, respectively. Both meteorites' spectra become bluer as grains sizes increase.\label{fig:grain}}
\end{figure*}

If the \jd{} variation is only due to a change in grain-sizes under a similar composition, we should be able to validate this by looking at the reflectance variation as a function of wavelength and grain-sizes. Figure \ref{fig:variation} bottom panel, shows that much of the \jd{} surface is spectrally red. We then considered whether different grain sizes on NIR spectral red meteorites could lead to negative-blue spectra. The effect of grain-sizes on iron-sulfide (Mundrabilla) and enstatite-achondrite (ALH78113) spectra are illustrated in Figure \ref{fig:grain}. The reflectance of the aubrite ALH78113 shows a flat NIR spectral slope for small grain-sizes ($<$45$\mu$m) and a drastic decline in the NIR slope as the grain-sizes increase up to chip-size grains. Similar behavior can be observed for the reflectance of Mundrabilla meteorite as a function of wavelength. The NIR slopes of ALH78113 ranges from -6\% to 0.6\% $\mu$m$^{-1}$, while the NIR slopes of Mundrabilla meteorite ranges from 26\% to 33\% $\mu$m$^{-1}$. In other words, the grain sizes of ALH78113 and Mundrabilla contributed a slope difference (redder slope minus bluest slope) of 6.6\% $\mu$m$^{-1}$ and 7\% $\mu$m$^{-1}$, respectively. \jd{} has a NIR slope difference of 41.71$\pm$0.01\% $\mu$m$^{-1}$ (most-positive to most-negative spectral slope difference). It is clear that grain sizes do have a strong influence on the NIR slope and visible color. However, we have demonstrated that is unlikely to explain the NIR slope variation of \jd{} with the size range of grains in RELAB. More laboratory work needs to be done on aubrites meteorites or simulant material at a larger range of grain to test the feasibility of the grain size hypothesis. Thus, the scenario where \jd{}'s spectral variation could be due to a compositional and grain sizes heterogeneity seems feasible.

\subsection{Color and Albedo: Hints for Mineralogical Composition \label{sec:Color}}

The colors of asteroids reflect information about their mineralogical composition since they can contain diagnostic features of the spectral bands. We developed an analysis to compare \jd{}'s visible colors and NIR slopes with meteorites and asteroids presented in Section \ref{sec:M-A}. This would help to explore the possible mineralogical composition of the spectral variability. Similarly, we provide a NIR spectral slope and albedo study to investigate possible compositional connections.

To obtain the Sloan $g^{\prime}$ - $r^{\prime}$ color, we retrieved the visible spectra of ALH78113, Mundrabilla troilite, Paragould troilite, troilite mineral, Peña Blanca silicate and sulfide samples, and Khor Temiki from RELAB. We convolved each spectrum with the bandpasses for Sloan $g^{\prime}$ and $r^{\prime}$ filters \citep{Gunn1998,Doi2010}. From the ratio of these integrated reflectances, we derived the intrinsic visible reflectance color, in magnitudes, for each meteorite. Finally, we add in the $g^{\prime}$ - $r^{\prime}$ Solar color of \citet{Willmer2018} to produce the apparent color that each meteorite would have if observed in space (that is, as illuminated by the Sun). The LCO intrinsic visible reflectance color of \jd{} was also determined using the same methodology. The LDT-LMI $g^{\prime} - r^{\prime}$ colors were selected based on the measured color at the corresponding time of the respective NIR spectra presented in Figure \ref{fig:XB}. To obtain the $g^{\prime}$ - $r^{\prime}$ color of the M- and E-types, we used their B-V colors given by \citet{TedescoUBV} and the color transformations between the Johnson-Cousins and Sloan system \citep{Castro}. 

\begin{figure*}[t!]
    \centering
    \includegraphics[width=0.9\textwidth]{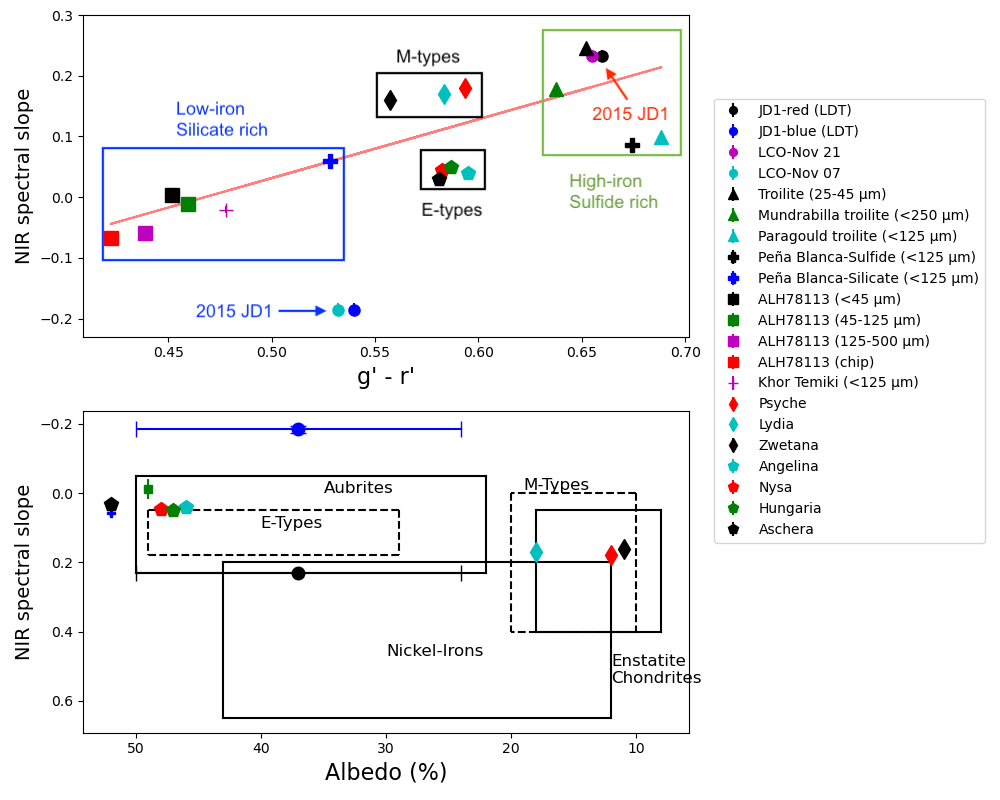}
    \caption{NIR spectral slope as a function of the $g^{\prime} - r^{\prime}$ Sloan colors (top panel) and geometric albedo (bottom panel) of \jd{} (filled circles), E-types (pentagons), M-types (diamonds), ALH78113 (squares), Peña Blanca aubrite (filled cross), Khor Temiki aubrite (thin cross), and minerals (triangles). Red line represent a linear relationship between NIR spectral slope (s$^{\prime}$) and Sloan $g^{\prime} - r^{\prime}$ color given by s$^{\prime}$ = ($g^{\prime} - r^{\prime}$) * (0.969 $\pm$ 0.291) - 0.453. \label{fig:ss_gr}}
\end{figure*}

Figure \ref{fig:ss_gr} top panel, illustrates the NIR spectral slope as a function of the $g^{\prime}$ - $r^{\prime}$ color of the selected RELAB meteorites, minerals, M- and E-types, and \jd{}'s colors. The NIR spectral slope was measured from 0.8 to 2.5 $\mu$m region. The solid circles in Figure \ref{fig:ss_gr} correspond to the observed colors of \jd{} from LDT and LCO. The black and blue circles represent the redder and bluer photometric color from LDT. We adopted the same NIR spectral slopes from the LDT spectroscopic observations to the retrieved colors from LCO visible spectra given that there is no NIR spectra from LCO. However, such assumption will not affect the purpose of this analysis because when comparing LCO and LDT we only refer to the visible colors. The retrieved LCO visible colors of \jd{} (magenta and cyan circles in Figure \ref{fig:ss_gr}) support a color variation, similar to the obtained from the photometric observation at LDT. The bottom panel, illustrates the NIR spectral slope as a function of albedo. The black boxes (aubrites, nickel-iron, and ensatite chondrite) indicates possible meteorites analog ranges from \citet{Gaffey1976}, and the dashed black boxes are the ranges derived from observation of M- and E-types from \citet{Bell1988}. The M- and E-types asteroids albedos were obtained from \citet{Masiero2014}, while aubrites Peña Blanca and ALH78113 from \citet{Fornasier2008}. We assumed the same albedo for the red and blue regions of \jd{}. A complete discussion of these results is provided in Section \ref{sec:composdis}.

\section{Dynamical Study \label{sec:dynamics}}

Pairing the spectroscopic analysis from Section \ref{sec:compositional analysis} to a dynamical study would provide constraints on \jd{}'s origin and possible connection to a parent body. We used the NEO dynamical model of \citet{Granvik2018} to calculate where \jd{} came from based on its current eccentricity, inclination, and semimajor axis of 0.222, 19.139$^{\circ}$, and 1.216 au, respectively.

According to the \citet{Granvik2018} model, which takes into account high-inclination orbits, \jd{} has an $84\pm2.1$\%  probability of coming from the $\nu_6$ secular resonance or other resonances close by, $11\pm1.2$\% chance of coming from the 3:1 mean-motion resonance (MMR) with Jupiter or other resonances close by, $5\pm0.8$\% chance of coming from the Hungaria group, and $0.4\pm0.1$\% chance of coming from the Phocaea group. Based on these probabilities, the most likely sources of origin of \jd{} would be regions in the asteroid belt feeding the $\nu_6$ complex or the 3:1 complex. 

In Section \ref{sec:compositional analysis}, we presented compositional connection between \jd{} and E[II] types or Angelina-like bodies. (64) Angelina's orbit (e,i,a) = (0.125, 1.309$^{\circ}$, 2.68 au) does not appear to agree with \jd{}'s high orbital inclination. However, the relatively low semimajor axis of \jd{} suggests that it has spent a relatively long time in the NEO region and has therefore also endured a non-negligible number of close planetary encounters. These planetary encounters may have changed \jd{}'s inclination dramatically. Hence, while (64) Angelina has not been linked as a possible parental body of E-type asteroids, we find it plausible that \jd{} may be genetically related to (64) Angelina, and have entered NEO region through the 3:1 MMR. Given its small size and hence rapid drift in semimajor axis caused by the Yarkovsky effect, it may also be possible that \jd{} has moved across the 3:1 MMR rapidly enough not to be ejected from the asteroid belt, and only been ejected when it reached the $\nu_6$ resonance \citep[see Fig.~17 in][]{2017A&A...598A..52G}. 

On the other hand, (434) Hungaria (e,i,a) = (0.073, 22.509$^{\circ}$, 1.94 au) has similar orbital parameters to that of \jd{} and the dynamical model suggests 5\% probability of coming from the Hungaria group. However, the compositional analysis of \jd{} does not agree with an origin in the Hungaria group as well as it does for Angelina-like bodies (see Sect.~\ref{sec:M-A}). 

Dynamical considerations suggest that both the $\nu_6$ resonance complex and the 3:1 resonance complex are potential escape routes for \jd{}. Assessing whether \jd{} reached the $\nu_6$ or the 3:1 resonance complexes after being ejected from an E-type family member (i.e., the Hungaria family or a non-Hungaria family), requires extensive dynamical modeling, so we leave it for future work. We do suggest, from compositional evidence, that \jd{} falls within the Angelina-like group \citet{Clark2004a} given the strong evidence for sulfide on its surface, which is not detected in the spectra of (434) Hungaria.

\section{Discussion \label{sec:discussion}}

\subsection{Circular polarization ratio and radar albedo}

The SC/OC ratio or CPR has commonly been used as a single-parameter study for surface composition and near-surface roughness \citep{Ostro1993, Benner2008}. Radar scattering modeling have manifest the non-linear behaviour of SC/OC ratios with composition and surface roughness mainly due to variable near-surface electric properties among different taxonomies \citep{Virkki2016}. Although attempts from both observational and modeling approaches have intended to understand the behaviour of the radar scattering properties of asteroids surfaces with respect its composition and particle size distribution, the area remain substantially unexplored and inconclusive. Future laboratory and modeling works would help to pivot our understanding regarding SC/OC ratio including the identification of near-surface elements controlling radar scattering properties.

\begin{figure}[ht!]
\includegraphics[width=0.47\textwidth]{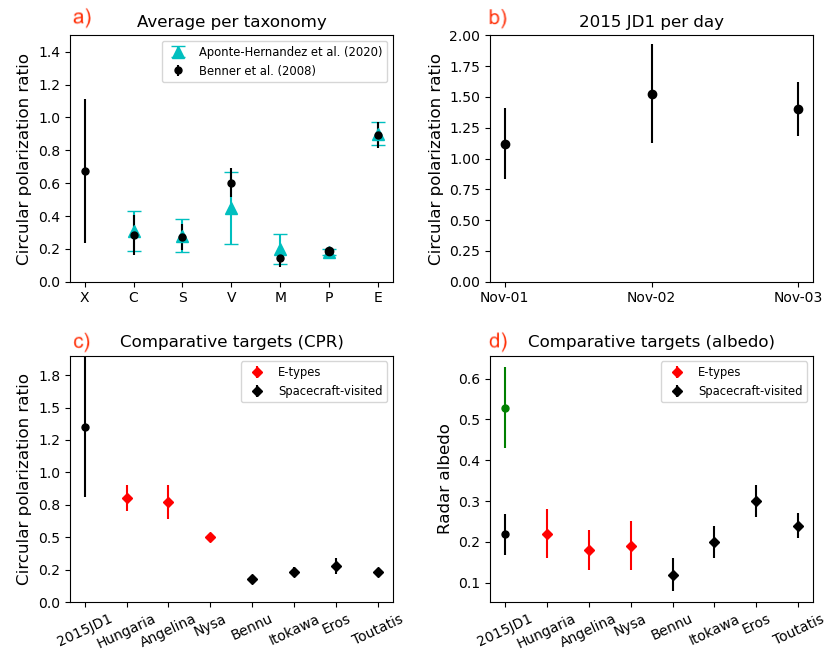}
\caption{Panel a: circular polarization ratios from \citet{Benner2008} and \citet{Aponte2020} as a function of the Bus-DeMeo taxonomy \citet{Demeo2009}. Panel b: circular polarization ratios of \jd{} as a function of observing days. Panel c: circular polarization ratios of \jd{} (black dot), selected E-types (red filled diamonds), and spacecraft-visited near-Earth asteroids (black filled diamonds). Panel d: comparative radar albedos for the asteroids in panel c. The black and green circles correspond to the OC and total radar albedo of \jd{}, respectively.  \label{fig:cpr-albedo}}
\end{figure}

\citet{Benner2008} published average CPR values for NEAs ranging from 0.143 $\pm$ 0.055 for M-type asteroids to 0.892 $\pm$ 0.079 for E-type asteroids. Recently, \citet{Aponte2020} also found similar trends among taxonomic classes. \citet{Aponte2020} found that asteroids with CPR $>$ 0.75 can be statistically, within 2-$\sigma$, characterized as E-types. E-types have presented robust statistical consistency on high CPR values \citep{Benner2008, Busch2008, Reddy700, Aponte2020}. Such a tendency might be illustrating characteristic properties of their mineralogical composition and surface properties. Based on this CPR review, we argue that \jd{}'s composition should be similar to those E-types asteroids (see Figure \ref{fig:cpr-albedo}). 
Although no spacecraft has visited an E-type asteroid to do a direct comparison with \jd{}, valuable information can be extracted from other spacecraft-visited asteroids that have radar observations. Comparatively, the CPR value of \jd{} exceeds all of those CPR of spacecraft-visited asteroids including S- and C-complex: (433) Eros (0.28 $\pm$ 0.06; \citet{Magri2001}), (25143) Itokawa (0.26 $\pm$ 0.04; \citet{Ostro2004}), (4179) Toutatis (0.23 $\pm$ 0.03; \citet{Magri2001}), and (101955) Bennu (0.18 $\pm$ 0.03; \citet{Nolan2013}). Given the uniqueness of the \jd{} scattering properties and lack of interpretation of such high CPR value, it it not possible to make connection with those spacecraft-observed asteroids (i.e, Bennu \citep{Lauretta2019}, Toutatis \citep{Jiang2015}, Eros \citep{Robinson2002}, and Itokawa \citep{Susorney2019}) surface properties.

Radar albedo can be used to infer surface properties (i.e., metallicity and density) of asteroids \citep{SHEPARD2008a,SHEPARD2010}. From radar data compilation \citep{Neese2012}, the average $\hat{\sigma}_{T}$ of asteroids is 15.7 $\pm$ 11.1\% (including near-Earth and main belt asteroids). The total power (SC + OC = $\hat{\sigma}_{T}$) of radar observed E-types asteroids (\textit{i.e.}, (64) Angelina \citep{Shepard2011}, (434) Hungaria \citep{Shepard2008c}, (44) Nysa \citep{Shepard2008c}, and 1998 $WT_{24}$ \citep{Busch2008}) range from 0.18 to 0.42, 0.24 to 0.32 for the M-type (16) Psyche \citep{SHEPARD2017}, and 0.12 to 0.3 for the spacecraft-visited asteroids shown in Figure \ref{fig:cpr-albedo} panel d. Comparatively, \jd{}'s total power ($\hat{\sigma}_{T}$=0.53$\pm$0.10) radar albedo exceeds all those reported for E-types. Moreover, \jd{}'s OC radar albedo is indistinguishable between the selected comparative objects, but higher than Bennu's total power.  Bennu has a radar albedo of 0.12 and a bulk density of 1.19 g/cm$^{3}$ \citep{Lauretta2019} and comets have typical radar albedo ranging from 0.4 to 0.1 and densities below 1 g/cm$^{3}$ \citep{Harmon1999, Harmon2004, Harmon2011}, thus we can reject a low near-surface density for \jd{}. 

The comparison between \jd{}'s disk-integrated radar albedo with those of E- and M-types, suggest some metal content near the surface. Adopting \citet{SHEPARD2010} relationship, we obtained an approximate upper-limit density of $2.5\ \mathrm{g}/\mathrm{cm}^3$ for a 30\% porosity using typical aubrites grain densities of $3.5\ \mathrm{g}/\mathrm{cm}^3$ from \citet{Britt2003}.

\subsection{Compositional Analysis of the Surface \label{sec:composdis}}

Figure \ref{fig:ss_gr} top panel has four boxes indicating either regions with similar composition or similar taxonomies, i.e., black boxes indicates E- and M-types, the green box shows meteorites and minerals rich in iron and sulfide, and the blue box indicates meteorites rich in enstatite silicate with low iron abundances. The first fact to notice is the mineralogical composition effect. This analysis shows that meteorites with higher iron and sulfide  abundances have redder spectral slope and $g^{\prime} - r^{\prime}$ color while bluer colors and spectral slopes are observed for meteorites whose mineralogical composition is shown to be richer in enstatite silicates and lower in iron. M-types have steeper spectral slopes than E-types due to the higher iron content, which can also be observed in our sample of well-known E$/$M-types asteroids. Second, \jd{}'s negative NIR spectral slope and blue colors deviate from the positive NIR spectral slope and red colors suggesting that such a blue region is compositionally distinct. \jd{}'s spectral-red surface has a steeper slope and redder Sloan colors than E$/$M-types and is mineralogically consistent with the spectral properties of iron sulfides (i.e., the mineral troilite and Peña Blanca sulfide region). Similarly, the red NIR spectral slope of \jd{} as a function of its geometric albedo place it between the aubrites and nickel-iron regions, suggesting as well some iron composition. The spectral-blue region has a bluer slope than the selected aubrites and than the aubrites slopes ranges presented in the bottom panel. The blue Sloan colors of \jd{} are similar to that of Peña Blanca silicate region. Such spectral similarities with aubrites ALH78113, Peña Blanca silicate region, and Khor Temiki suggest a composition abundant in enstatite-silicate and low iron content. The blue region could still preserve some sulfide end-members since both LCO spectra shows the 0.5 $\mu$m feature.

In Figure \ref{fig:M&A} left panel, two spectra of Peña Blanca meteorite are shown. The steeper reflectance (red dashed line) is from the sulfide-rich region while the other one is from an enstatite-silicate region. The NIR similarity between \jd{} red-sloped spectra and Peña Blanca-sulfide region suggest a common mineralogical composition, while \jd{}'s blue Sloan colors support a Peña Blanca silicate-like composition. Peña Blanca Spring is a fragmental breccia of particular interest since it contains large crystals of enstatite ($\sim$10 cm in diameter from the largest extension), which has been attributed to an igneous formation \citep{Lonsdale}. The bulk analysis of Peña Blanca Spring, which correspond to the silicate-region spectra referred in this work, revealed a high abundance of enstatite (95 wt.\%) with low iron, plagioclase (2.1 wt.\%), diopside (2.7 wt.\%), forsterite (0.3 wt.\%) and sulfide end-members in lower abundances (0.2 wt.\%) \citep{Watters1979, Lonsdale}. The chemical analysis of the Peña Blanca sulfide region mentioned here, yielded a high iron abundance and a Fe:Ni of 22.68 \citep{Lodders1993}. This comparison between Peña Blanca chemical analysis and \jd{} is in agreement with the analysis of NIR spectral slope and Sloan colors.

ALH78113 is another fragmental breccia with chondritic-like inclusions, similar to Cumberland Falls, with large enstatite fragments ($\sim$5 cm in diameter). Based on those inclusions, \citet{Lipschutz1988} concluded that ALH78113 is a representative of foreign primitive fragments, not necessarily formed under the same conditions as enstatite achondrites meteorites, that collided with the aubrite parent body. The bulk chemical analysis indicate that enstatite is the most abundant mineral \citep{Kimura1993}. Their silicate phases shows feldspar, olivine, glass, diopside, enstatite, and silica mineral \citep{Lipschutz1988, Kimura1993}. The bluest spectral slope of \jd{} is $\sim$10\% more negative than the bluer spectral slope of ALH78113 (see the green spectrum with grains $>$chip-size in Figure \ref{fig:grain} left panel). It is likely, as indicated by the comparison with Peña Blanca and from the slope-color analysis, that the blue (spectral) region on \jd{} contains lower abundances of Fe/Ni and larger grain sizes than ALH78113 and than the rest of its surface.

\subsection{Possible explanations for the spectral variability}

In Section \ref{sec:composdis}, we investigated the spectral slope and color variation of \jd{}. However, it is difficult to explain the nature of \jd{}'s spectral variability with the available data. Therefore, we provide the following hypotheses that could yield the observed variability.

\begin{enumerate}
    
    \item \textit{Space weathering}: Several active comets are known to show longitudinal spectral variations in their surfaces in a relatively limited time-frame \citep{Lukyanyk2019} due to the sublimation of icy grains. The NIR spectra of (6478)~Gault, which is known to be an active asteroid, showed a slope variation of 22.6\% $\mu$m$^{-1}$ \citep{Marsset_2019}. They hypothesized that the NIR spectral variation was likely due to a loss of regolith, resulting in the exposure of newer material. As mentioned in section \ref{sec:dynamics}, the low semimajor axis of \jd{} suggest considerable number of close planetary encounters. It could be possible that these planetary encounters played a role resurfacing the body (i.e., near its neck) yielding the exposure of unweathered or fresher (bluer) material.

    \item \textit{Accretion}: \jd{} could have been formed through the accretion of heterogeneous fragments. This could be the case where the parent body of \jd{} was disrupted by a compositionally different impactor or the progenitor itself possessed heterogeneity.

    \item \textit{Exogenous material}: The recent spacecraft-visited asteroids Ryugu and Bennu revealed bright boulders spectrally distinct from the dark ones on their surfaces. Recent studies suggest that such bright boulders represent remnants of exogenous compositionally different material \citep{Tatsumi2021, Dellaguistina2021}. In the case of \jd{}, partial surface blanketing by exogenous material could have created a spectrally blue surface `patch' covering a notable portion of its surface. In addition, our compositional analysis shows spectral similarities between \jd{}'s blue spectra and the aubrite ALH781113. \citet{Lipschutz1988} suggested that ALH78113 could have been a surviving fragment from an external igneous body that collided with the E-type progenitor/s. Furthermore, the exogenous material on Bennu and Ryugu made up a tiny fraction of their surfaces (<0.1\%). While the likelihood of having a significant portion of JD1’s surface contaminated by exogenous material seems very low, this hypothesis may still be plausible.

\end{enumerate}

\section{Conclusions \label{sec:conclusions}}

Through the combination of ground-based telescopic observations covering the visible, near-infrared, and radar wavelengths, we physically characterized the NEA \jdlong{}, finding interesting and possibly unique surface features of this small airless body. In summary, we found that \jd{} has a contact binary nature with a head of $\sim$50~m and a body of $\sim100$~m across the longest visible axis projection, or approximately between 150 and 190~m in diameter. \jd{} has a high circular-polarization ratio between 1.12 to 1.52, one of the highest values measured for NEAs, and also a high total radar albedo of 0.53$\pm$0.01. We measured a high geometric albedo of $0.35\pm0.12$ that is consistent with an E-type taxonomy \citep{Thomas2011}. We derived a g$^{\prime}$ - r$^{\prime}$ color of 0.6 $\pm$ 0.03. Our dynamical study suggests that both the $\nu_6$ resonance complex and the 3:1 resonance complex are potential escape routes for \jd{}. Spectral comparison suggest a compositional link between \jd{} and the Angelina-like group from \citet{Clark2004a}.

Our rotationally-resolved spectroscopy revealed red and blue NIR spectra in the surface of \jd{}. Our comprehensive analysis of the observed rotational spectral variability allows us to draw the following conclusion: \jd{} is likely an E-type asteroid rich in sulfides and iron with a portion of its surface that could consist of larger grain sizes and/or mineralogically richer in silicates and deficient in iron, responsible for the spectral variability observed. \jd{} could be one of the first sub-km NEA to show spectral variability likely associated with an heterogeneous composition and grain sizes.

We plan to continue this investigation in order to validate spectral heterogeneity in the surface of \jd{} using spectrophotometry and spectroscopy observations. The next close approach of \jd{} will take place around 2023 April with a predicted $V$-band magnitude of 19.8. The next radar apparition at a comparable distance to that in 2019 ($0.034~au$) will occur near 2058 November when \jd{} is expected to have a $V$ magnitude of 15.1.  \\

\acknowledgements 
This work was supported in part by the Arizona Board of 
Regents,
Arizona's Technology and Research Initiative Fund, and by NASA grant No.\ NNX15AF81G to DET.
These results 
made use of the Lowell Discovery Telescope (LDT) at Lowell
Observatory. LDT is localized on homelands sacred to Native Americans. We thank them for allowing part of their sacred lands to be used for astronomical research. Lowell is a private, non-profit institution
dedicated to astrophysical research and public appreciation
of astronomy and operates the LDT in partnership with Boston
University, the University of Maryland, the University of
Toledo, Northern Arizona University and Yale University.

The Near-Infrared High Throughput Spectrograph was funded
by NASA award No.\ NNX09AB54G through its Planetary Astronomy
and Planetary Major Equipment programs. The Large Monolithic
Imager was built by Lowell Observatory using funds provided
by the National Science Foundation (AST-1005313). 

\ Part of 
this work was done at the Arecibo Observatory, which is
a facility of the National Science Foundation operated 
under cooperative agreement by the University of Central
Florida, Yang Enterprises, Inc., and Universidad Metropolitana.
The Arecibo Planetary Radar Program is supported by NASA 
grant 80NSSC19K0523 and NNX13AQ46G through the Near-Earth Object Observations
program. 

\ This material is in part supported by the National
Science Foundation Graduate Research Fellowship Program under
grant No. 2018258765 to COC. Any opinions, findings, and conclusions
or recommendations expressed in this material are those of
the author(s) and do not necessarily reflect the views of 
the National Science Foundation. 

Part of the computational
analyses were carried out on Northern Arizona University's
Monsoon computing cluster, funded by Arizona's Technology 
and Research Initiative Fund. This research has made use 
of data and/or services provided by the International
Astronomical Union's Minor Planet Center. This research
has made use of NASA's Astrophysics Data System. This 
research has made use of the SkyBoT Virtual Observatory tool
\citep{Berthier:2006tn}. 

\facility{Arecibo, Lowell Discovery Telescope, Las Cumbres Observatory}

\software{\textit{Spextool} \citep{Cushing}, \textit{Astropy} \citep{AstroPy2013},
          \textit{Scipy} \citep{scipy2020},
          \textit{Numpy} \citep{vanderWalt2011,numpy2020},
          \textit{Matplotlib} \citep{matplotlib2007}, 
          \textit{Photometry Pipeline} \citep{Mommert},
          SExtractor \citep{1996A&AS..117..393B}, 
          SCAMP \citep{Bertin}, 
          SAO Image DS9 \citep{2003ASPC..295..489J}, Aperture Photometry Tool \citep{Laher2012}, Horizons System\footnote{\url{https://ssd.jpl.nasa.gov/horizons/app.html}}}


\bibliography{referencias}{}
\bibliographystyle{aasjournal}

\end{document}